\documentclass[preprint,12pt,3p]{elsarticle}

\usepackage{amssymb}
\usepackage{amsthm}
\usepackage{amsmath}
\usepackage{amsfonts}
\usepackage{graphics}
\usepackage{graphicx}

\begin{document}

\begin{frontmatter}

\title{Ordered Kripke Model, Permissibility, and Convergence of
Probabilistic Kripke Model\tnoteref{label0}}
\tnotetext[label0]{The author would like to thank Andr\'{e}s Perea and Zsombor Z. M\'{e}der for
their valuable comments and encouragements. She gratefully acknowledge the
support of Grant-in-Aids for Young Scientists (B) of JSPS No. 17K13707,
Grant for Special Research Project No. 2017K-016 of Waseda University. }

\author{Shuige Liu}
\address{Faculty of Political Science and Economics, Waseda University, 1-6-1
Nishi-Waseda, Shinjuku-Ku, 169-8050, Tokyo, Japan}

\ead{shuige\_liu@aoni.waseda.jp}

\begin{abstract}
We define a modification of the standard Kripke model, called the ordered
Kripke model, by introducing a linear order on the set of accessible states
of each state. We first show this model can be used to describe the
lexicographic belief hierarchy in epistemic game theory, and perfect
rationalizability can be characterized within this model. Then we show that
each ordered Kripke model is the limit of a sequence of standard
probabilistic Kripke models with a modified (common) belief operator, in the
senses of structure and the ($\varepsilon $-)permissibilities characterized
within them.
\end{abstract}


\end{frontmatter}

\section{Preliminaries\label{sec:suy}}

In this section we give surveys on lexicographic belief and permissibility
(Section \ref{sec:suy}.1) and on probabilistic Kripke model for games
(Section \ref{sec:suy}.2). These will be preparation for the introduction of
ordered Kripke model in Section \ref{sec:las}.

\subsection{Lexicographic belief and permissibility}

In this subsection we give a survey on lexicographic epistemic model (with
complete information) and the definition of permissibility. For a details,
see Perea \cite{p12}, Chapter 5. Consider a finite 2-person strategic form
game $G=(N,\{S_{i}\}_{i\in N},\{u_{i}\}_{i\in N})$ where $I=1,2$. A finite 
\emph{lexicographic epistemic model} for $G$ is a tuple $M^{lex}=(\Theta
_{i},\beta _{i})_{i\in N}$ where\smallskip \newline
(a) $\Theta _{i}$ is a finite set of types, and\smallskip \newline
(b) $\beta _{i}$ is a mapping that assigns to every $\theta _{i}\in \Theta
_{i}$ a lexicographic belief over $\Delta (S_{j}\times \Theta _{j}),$ i.e., $%
\beta _{i}(\theta _{i})=(\beta _{i1},\beta _{i2},...,\beta _{iK})$ where $%
\beta _{ik}\in \Delta (S_{j}\times \Theta _{j})$ for $k=1,...,K.$\smallskip 

Let $\theta _{i}\in \Theta _{i}$ with $\beta _{i}(\theta _{i})=(\beta
_{i1},\beta _{i2},...,\beta _{iK}).$ Each $\beta _{ik}$ $(k=1,...,K)$ is
called $\theta _{i}$'s \emph{level-}$k$ belief. For $(s_{j},\theta _{j})\in
S_{j}\times \Theta _{j},$ we say $\theta _{i}$ \emph{deems} $(s_{j},\theta
_{j})$ \emph{possible} iff $\beta _{ik}(s_{j},\theta _{j})>0$ for some $k\in
\{1,...,K\}.$ We say $\theta _{i}$ \emph{deems} $\theta _{j}\in \Theta _{j}$ 
\emph{possible} iff $\theta _{i}$ deems $(s_{j},\theta _{j})$ possible for
some $s_{j}\in S_{j}$. For each $\theta _{i}\in \Theta _{i},$ we denote by $%
\Theta _{j}(\theta _{i})$ the set of all $\theta _{j}\in \Theta _{j}$ deemed
possible by $\theta _{i}$.\smallskip \newline
\textbf{Definition \ref{sec:suy}.1 (Caution)} Type $\theta _{i}\in \Theta
_{i}$ is \emph{cautious} iff for each $\theta _{j}\in \Theta _{j}(\theta
_{i})$ and each $s_{j}\in S_{j},$ it deems $(s_{j},\theta _{j})$
possible.\smallskip 

For each $s_{i}\in S_{i}$, let $u_{i}(s_{i},\theta _{i})=(u_{i}(s_{i},\theta
_{i1}).,..,u_{i}(s_{i},\theta _{iK}))$ where for each $k=1,...,K,$ $%
u_{i}(s_{i},\theta _{ik}):=\Sigma _{(c_{j},t_{j})\in C_{j}\times T_{j}}\beta
_{ik}(s_{j},\theta _{j})u_{i}(s_{i},s_{j}),$ that is, each $%
u_{i}(s_{i},\theta _{ik})$ is the expected utility for $s_{i}$ over $\theta
_{ik}$ and $u_{i}(s_{i},\theta _{i})$ is a vector of expected utilities. For
each $s_{i},s_{i}^{\prime }\in S_{i}$, we say that $\theta _{i}$ \emph{%
prefers} $s_{i}$ \emph{to} $s_{i}^{\prime }$, denoted by $u_{i}(s_{i},\theta
_{i})>u_{i}(s_{i}^{\prime },\theta _{i}),$ iff there is $k\in \{0,...,K-1\}$
such that the following two conditions are satisfied:\smallskip \newline
(a) $u_{i}(s_{i},\theta _{i\ell })=u_{i}(s_{i}^{\prime },\theta _{i\ell })$
for $\ell =0,...,k,$ and\smallskip \newline
(b) $u_{i}(s_{i},\theta _{i,k+1})>u_{i}(s_{i}^{\prime },\theta _{i,k+1})$%
.\smallskip \newline
We say that $\theta _{i}$ \emph{is indifferent between }$s_{i}$ \emph{and }$%
s_{i}^{\prime },$ denoted by $u_{i}(s_{i},\theta _{i})=u_{i}(s_{i}^{\prime
},\theta _{i}),$ iff $u_{i}(s_{i},\theta _{ik})=u_{i}(s_{i}^{\prime },\theta
_{ik})$ for each $k=1,...,K.$ It can be seen that the preference relation on 
$S_{i}$ under each type $\theta _{i}$ is a linear order. $s_{i}$ is \emph{%
rational} (or \emph{optimal}) for $\theta _{i}$ iff $\theta _{i}$ does not
prefer any choice to $s_{i}$.\smallskip \newline
\textbf{Definition \ref{sec:suy}.2 (Primary belief in the opponent's
rationality) }Let $\theta _{i}\in \Theta _{i}$ with $\beta _{i}(\theta
_{i})=(\beta _{i1},\beta _{i2},...,\beta _{iK}).$ $\theta _{i}$ \emph{%
primarily believes in }$\emph{j}$\emph{'s rationality} iff $\theta _{i}$'s
primary belief $\theta _{i1}$ only assigns positive probability to those $%
(s_{j},\theta _{j})$ where $s_{j}$ is rational for $\theta _{j}.$\smallskip 
\newline
\textbf{Definition \ref{sec:suy}.3 (Common full belief in a property) }Let $P
$ be an arbitrary property of lexicographic types.\smallskip \newline
(a) $\theta _{i}\in \Theta _{i}$ \emph{expresses }$0$\emph{-fold full belief
in} $P$ iff $\theta _{i}$ satisfies $P;$\smallskip \newline
(b) For each $n\in \mathbb{N},$ $\theta _{i}\in \Theta _{i}$ \emph{expresses 
}$(n+1)$\emph{-fold full belief in} $P$ iff $\theta _{i}$ only deems
possible $j$'s types that express $n$-fold full belief in $P.$\smallskip 
\newline
$\theta _{i}$ \emph{expresses common full belief in} $P$ iff it expresses $n$%
-fold full belief in $P$ for each $n\in \mathbb{N}.$\smallskip \newline
\textbf{Definition \ref{sec:suy}.4 (Permissibility)}. Given a lexicographic
epistemic model $M^{lex}=(\Theta _{i},\beta _{i})_{i\in N}$ for a game $%
G=(N,\{S_{i}\}_{i\in N},\{u_{i}\}_{i\in N})$, $s_{i}\in S_{i}$ is \emph{%
permissible} iff it is optimal to some $\theta _{i}\in \Theta _{i}$ which
expresses common full belief in caution and primary belief in
rationality.\smallskip \newline
\textbf{Example \ref{sec:suy}.1 }Consider the game $G$ as follows (Myerson 
\cite{m78}):%
\begin{equation*}
\begin{tabular}{|l|l|l|}
\hline
$u_{1}\backslash u_{2}$ & $C$ & $D$ \\ \hline
$A$ & $1,1$ & $0,0$ \\ \hline
$B$ & $0,0$ & $0,0$ \\ \hline
\end{tabular}%
\end{equation*}%
and $M^{lex}=(\Theta _{i},\beta _{i})_{i\in N}$ for $G$ where $\Theta
_{1}=\{\theta _{1}\},$ $\Theta _{2}=\{\theta _{2}\},$ and%
\begin{equation*}
\beta _{1}(\theta _{1})=((C,\theta _{2}),(D,\theta _{2})),\text{ }%
b_{2}(t_{2})=((A,\theta _{1}),(B,\theta _{1})).
\end{equation*}%
It can be seen that $A$ is permissible since it is optimal to $t_{1}$ which
expresses common full belief in caution and primary belief in
rationality.\smallskip 

It is shown by Proposition 5.2 in Asheim and Dufwenberg \cite{ad03} that a
strategy is permissible if and only if it survives an algorithm called \emph{%
Dekel-Fudenberg procedure }(Dekel and Fudenberg \cite{df90}). Given a game $%
G,$ by Dekel-Fudenberg procedure we mean the process that (1) at first round
we eliminate all weakly dominated strategies in $G,$ and (2) then iteratedly
eliminate dominated strategies until no strategies can be eliminated.

\subsection{Probabilistic Kripke model for games}

In this subsection we give a survey of the probabilistic Kripke model for
games which is a generalization of the standard Kripke model that is able to
capture both pure and mixed strategies. For details, see Bonanno \cite{b08}, 
\cite{b15}. Let $G=(N,\{S_{i}\}_{i\in N},\{u_{i}\}_{i\in N})$ be a 2-person
strategic form game. A \emph{probabilistic Kripke model of }$G$ is a tuple $%
\mathcal{M}=(W,\{R_{i}\}_{i\in N},\{p_{i}\}_{i\in N},\{\sigma _{i}\}_{i\in
N})$ where \smallskip \newline
(1) $W\neq \emptyset $ is the set of \emph{states} (or \emph{possible worlds}%
), sometimes called the \emph{domain} of $\mathcal{M}$ and is denoted by $%
\mathcal{D}(\mathcal{M});$\smallskip \newline
(2) For each $i\in N$, $R_{i}\subseteq S\times S$ is the \emph{accessibility
relation} for player $i.$ For each $w\in W,$ we use $R_{i}(w)$ to denote the
set of all accessible states from $w,$ i.e., $R_{i}(w)=\{w^{\prime }\in
W:wR_{i}w^{\prime }\};$\smallskip \newline
(3) For each $i\in N,$ $p_{i}$ is a mapping from $W$ to $\Delta (W)$
satisfying (a) for each $w\in W,$ supp $p_{i}(w)\subseteq R_{i}(w),$ and (b)
for each $w^{\prime }\in R_{i}(w),$ $p_{i}(w^{\prime })=p_{i}(w);$\smallskip 
\newline
(4) For each $i\in N,$ $\sigma _{i}$ is a mapping from $W$ to $S_{i}$ such
that for each $w^{\prime }\in R_{i}(w),$ $\sigma _{i}(w^{\prime })=\sigma
_{i}(w)$\textsf{.}\smallskip 

$(W,\{R_{i}\}_{i\in N},\{\sigma _{i}\}_{i\in N})$ is a \emph{standard Kripke
model of }$G$. $\mathcal{M}^{o}=(W,\{R_{i}\}_{i\in N})$ is called the \emph{%
Kripke frame} of $\mathcal{M}.$ Here we follow the literatures and assume
that $\mathcal{M}^{o}$ is a KD45 frame, i.e., each $R_{i}$ is serial,
transitive, and Euclidean. For each $i\in N,$ a \emph{semantic belief
operator} is a function $\mathbb{B}_{i}:2^{W}\rightarrow 2^{W}$ such that
for each $E\subseteq W,$%
\begin{equation}
\mathbb{B}_{i}(E)=\{w\in W:R_{i}(w)\subseteq E\}.  \label{sbo}
\end{equation}%
A \emph{semantic common belief operator} is a function $\mathbb{CB}%
:2^{W}\rightarrow 2^{W}$ such that for each $E\subseteq W,$%
\begin{equation}
\mathbb{CB}(E)=\{w\in W:\cup _{i\in N}R_{i}(w)\subseteq E\}.  \label{scbo}
\end{equation}%
It can be seen that $\mathbb{B}_{i}$ and $\mathbb{CB}$ correspond to Aumann 
\cite{a76}'s standard concept \textquotedblleft knowledge\textquotedblright\
and \textquotedblleft common knowledge\textquotedblright .

At $w\in W$ the strategy $s_{i}\in S_{i}$ with $s_{i}$ is \emph{at least as
prefered to} $s_{i}^{\prime }$ iff $u_{i}(s_{i},\Sigma _{w^{\prime }\in
R_{i}(w)}p_{i}(w)(w^{\prime })\sigma _{j}(w^{\prime }))\geq
u_{i}(s_{i}^{\prime },\Sigma _{w^{\prime }\in R_{i}(w)}p_{i}(w)(w^{\prime
})\sigma _{j}(w^{\prime })).$ $s_{i}$ is \emph{prefered to} $s_{i}^{\prime }$
at $w$ iff the strict inequality holds. $s_{i}$ is \emph{optimal }at $w$ iff
there is no strategy preferred to $s_{i}$ at $w$. A state $w$ is \emph{%
rational} for $i$ iff $\sigma _{i}(w)$ is optimal at $w$. We use $RAT_{i}$
to denote the set of all rational states for player $i,$ and define $%
RAT=\cap _{i\in N}RAT_{i}$.

The following statement connects iterated elimination of pure dominated
strategies (an algorithm) to rationality (an epistemic concept). Its proof
can be found in Bonanno \cite{b15}, p.452.\smallskip \newline
\textbf{Theorem \ref{sec:suy}.1 (Iterated elimination of dominated
strategies and Kripke model)}. Let $G=(N,\{S_{i}\}_{i\in N},\{u_{i}\}_{i\in
N})$ and $S^{IEDS}$ be the set of strategy profiles surviving iterated
elimination of dominated strategies. Then\smallskip \newline
(1) given an arbitrary probabilistic Kripke model of $G$, if $w\in \mathbb{CB%
}(RAT),$ then $\sigma (w)\in S^{IEDS};$\smallskip \newline
(2) for each $s\in S^{IEDS}$, there is a probabilistic Kripke model of $G$
and a state $w$ such that $\sigma (w)=s$ and $w\in \mathbb{CB}(RAT)$.

\section{Ordered Kripke Model of Games and Permissibility\label{sec:las}}

In this section we define the ordered Kripke model as a modification of the
standard one and show how it can be used to describe the lexicographic
reasoning in game theory.\smallskip \newline
\textbf{Definition \ref{sec:las}.1 (Ordered epistemic model)} Let $%
G=(N,\{S_{i}\}_{i\in N},\{u_{i}\}_{i\in N})$ be a 2-person strategic form
game. An \emph{ordered Kripke model} of $G$ is a tuple $\overline{\mathcal{M}%
}=(W,\{R_{i}\}_{i\in N},\{\sigma _{i}\}_{i\in N},\{\lambda _{i}\}_{i\in N})$
where\smallskip \newline
(1) $(W,\{R_{i}\}_{i\in N},\{\sigma _{i}\}_{i\in N})$ is a standard Kripke
model of $G$, and\smallskip \newline
(2) For each $i\in N$, $\lambda _{i}$ assigns to each $w\in W$ an injection
from a cut $\{1,...,K\}$ of natural numbers to the set of probability
distributions (with finite supports) over $R_{i}(w)$, i.e., $\lambda
_{i}(w):\{1,...,K\}\rightarrow \Delta (R_{i}(w)).$ $\lambda _{i}(w)$ can be
interpreted as a linear order on a finite subset of $\Delta (R_{i}(w)).$ We
use $\mathcal{D}(\lambda _{i}(w))$ and $\mathcal{R}(\lambda _{i}(w))$ to
denote the domain and the range of $\lambda _{i}(w)$, i.e., $\mathcal{D}%
(\lambda _{i}(w))=\{1,...,K\}$ and $\mathcal{R}(\lambda _{i}(w))=\{\lambda
_{i}(w)(1),...,\lambda _{i}(w)(K)\}$\textsf{.}\smallskip \newline
\textbf{Definition \ref{sec:las}.2 (Caution)}. Let $G=(N,\{S_{i}\}_{i\in
N},\{u_{i}\}_{i\in N})$ be a strategic form game and $\overline{\mathcal{M}}%
=(W,\{R_{i}\}_{i\in N},\{\sigma _{i}\}_{i\in N},\{\lambda _{i}\}_{i\in N})$
an ordered Kripke model for $G.$ $R_{i}$ is \emph{cautious} at $w\in W$ iff
for any $s_{j}\in S_{j}$ ($j\neq i$), there exists $w^{\prime }$ which is
assigned a possitive probability by some element in $\mathcal{R}(\lambda
_{i}(w))$ such that $\sigma _{j}(w^{\prime })=s_{j}$. We say $\overline{%
\mathcal{M}}$ is \emph{cautious} iff for each $i\in N$, $R_{i}$ is cautious
at every $w\in W$.\smallskip 

The difference between the ordered Kripke model and the standard one is that
the former assigns a linear order $\lambda _{i}(w)$ on $R_{i}(w)$ for each
state $w.$ This order is used to define the preferences in the model. We
have the following defintion.\smallskip \newline
\textbf{Definition \ref{sec:las}.3 (Lexicographic preferences) }Let $%
G=(N,\{S_{i}\}_{i\in N},\{u_{i}\}_{i\in N})$ be a strategic form game and $%
\overline{\mathcal{M}}=(W,\{R_{i}\}_{i\in N},\{\sigma _{i}\}_{i\in
N},\{\lambda _{i}\}_{i\in N})$ an ordered Kripke model for $G.$ At $w\in W$
the strategy $s_{i}\in S_{i}$ is \emph{at least as lexicographically
prefered to} $s_{i}^{\prime }$, denoted by $s_{i}\succeq _{w}s_{i}^{\prime },
$ iff $\exists k\in \{0,...,|\mathcal{D}(\lambda _{i}(w))|\}$ such
that\smallskip \newline
(a) $u_{i}(s_{i},\sigma _{j}(\sigma _{j}(\lambda
_{i}(w)(t))))=u_{i}(s_{i}^{\prime },\sigma _{i}(\lambda _{i}(w)(t)))$ for
all $t\leq k$;\smallskip \newline
(b) $u_{i}(s_{i},\sigma _{j}(\lambda _{i}(w)(k+1)))=u_{i}(s_{i}^{\prime
},\sigma _{i}(\lambda _{i}(w)(k+1))).$\smallskip 

Here by $\sigma _{j}(\lambda _{i}(w)(t))$ we mean the mixture of stategies
in $\sigma _{j}(\lambda _{i}(w)(t)).$ Therefore 
\begin{equation*}
u_{i}(s_{i},\sigma _{j}(\sigma _{j}(\lambda _{i}(w)(t))))=\Sigma _{w^{\prime
}\in R_{i}(w)}\lambda _{i}(w)(t)(w^{\prime })u_{i}(s_{i},\sigma
_{i}(w^{\prime })).
\end{equation*}%
It can be seen that when $k=|\mathcal{D}(\lambda _{i}(w))|,$ $s_{i}$ and $%
s_{i}^{\prime }$ generates the same payoff for player $i$ along $\lambda
_{i}(w).$ This case is denoted by $s_{i}\simeq _{w}s_{i}^{\prime }.$ When $%
k\neq |\mathcal{D}(\lambda _{i}(w))|,$ we say that $s_{i}$ is \emph{%
lexicographically prefered to} $s_{i}^{\prime }$ at $w,$ denoted by $%
s_{i}\succ _{w}s_{i}^{\prime }.$ $s_{i}$ is \emph{optimal} at $w$ iff there
is no $s_{i}^{\prime }\in S_{i}$ such that $s_{i}^{\prime }\succ _{w}s_{i}.$
We say a state $w$ is \emph{lexicographically rational }for $i$ iff the
choice $\sigma _{i}(w)$ is optimal for $i.$ For each $i\in N,$ let $LRAT_{i}$
be the set of rational states for player $i$ and $LRAT=\cap _{i\in N}LRAT_{i}
$.\smallskip \newline
\textbf{Example \ref{sec:las}.1}. Consider the following game $G$ in Example %
\ref{sec:suy}.1:%
\begin{equation*}
\begin{tabular}{|l|l|l|}
\hline
$u_{1}\backslash u_{2}$ & $C$ & $D$ \\ \hline
$A$ & $1,1$ & $0,0$ \\ \hline
$B$ & $0,0$ & $0,0$ \\ \hline
\end{tabular}%
\end{equation*}%
and an ordered Kripke model $\overline{\mathcal{M}}$ as follows:

\begin{figure}[h]
\centering\includegraphics[width=0.9\linewidth]{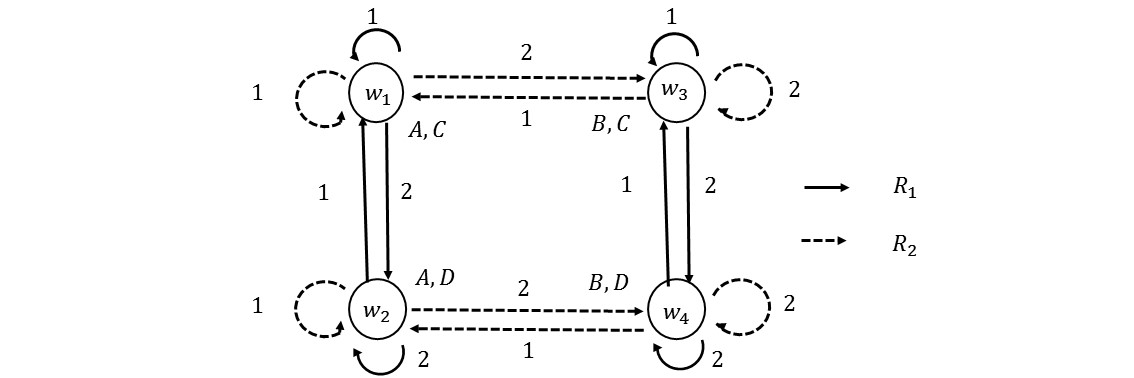}
\caption{An ordered Kripke model for $G$}
\end{figure}

It can be seen that $\overline{%
\mathcal{M}}$ is cautious. It can be seen that $A$ and $C$ are optimal in
each state, $w_{1}$ and $w_{2}$ are rational for player 1, and $w_{1}$ and $%
w_{3}$ are rational for player 2. Therefore, $LRAT_{1}=\{w_{1},w_{2}\}$, $%
LRAT_{2}=\{w_{1},w_{3}\},$ and $LRAT=\{w_{1}\}.$ On the other hand, as
mentioned in Example \ref{sec:suy}.1, since both $\sigma _{1}(w_{2})=A$ and $%
\sigma _{2}(w_{2})=D$ are permissible strategies, lexicographic rationality
in the ordered Kripke model here captures the concept of \textquotedblleft a
strategy is rational under a lexicographic belief\textquotedblright\ in the
first order. Now the problem is how to define belief hierarchy and common
belief in this model. It can be seen that we cannot adopt $\mathbb{B}_{i}$
and $\mathbb{CB}$ in standard approach. Indeed, in Example \ref{sec:las}.1 $%
\mathbb{B}_{i}(LART)=\mathbb{CB}(LART)=\emptyset ,$ which is contradictory
to our intention to preserve $w_{2}$. Here we give one approach. For each $%
i\in N$ and $w\in W,$ let $R_{i}^{1}(w)=\{w^{\prime }\in W:\lambda
_{i}(w)(1)(w^{\prime })>0\}$ and $R^{1}=\cup _{i\in N}R_{i}^{1}.$ A \emph{%
semantic level-1 belief operator }for player $i$ is a mapping $\mathbb{B}%
_{i}^{1}:2^{W}\rightarrow 2^{W}$ such that for each $E\subseteq W,$%
\begin{equation}
\mathbb{B}_{i}^{1}(E)=\{w\in W:R_{i}^{1}(w)\subseteq E\}.  \label{ism}
\end{equation}%
Similarly, a \emph{semantic common level-1 belief} operator is a mapping $%
\mathbb{CB}^{1}:2^{W}\rightarrow 2^{W}$ such that for each $E\subseteq W,$%
\begin{equation}
\mathbb{CB}^{1}(E)=\{w\in W:\cup _{i\in N}R_{i}^{1}(w)\subseteq E\}.
\label{cb}
\end{equation}%
It can be seen that $\mathbb{B}_{i}^{1}(LRAT)=\mathbb{CB}^{1}(LRAT)=\{w_{1}\}
$ in Example \ref{sec:las}.1. In general, we have the follwong
result.\smallskip \newline
\textbf{Theorem \ref{sec:las}.1 (Permissibility and semantic common level-1
belief)}. Let $G=(N,\{S_{i}\}_{i\in N},\{u_{i}\}_{i\in N})$ be a strategic
form game and $S^{PER}\subseteq S$ be the set of permissible strategy
profiles. Then\smallskip \newline
\textbf{(1)} given an arbitrary cautious ordered Kripke model of $G,$ if $%
w\in \mathbb{CB}^{1}(LRAT),$ then $\sigma (w)\in S^{PER}$, and\smallskip 
\newline
\textbf{(2)} for each $s\in S^{PER},$ there exists a cautious ordered Kripke
model of $G$ such that $\sigma (w)=s$ and $w\in \mathbb{CB}^{1}(LRAT).$%
\smallskip 

To show Theorem \ref{sec:las}.1, we need the following lemma.\smallskip 
\newline
\textbf{Lemma \ref{sec:las}.1}. Let $G=(N,\{S_{i}\}_{i\in N},\{u_{i}\}_{i\in
N})$ be a strategic form game and $S^{DF}\subseteq S$ be the set of strategy
profiles surviving Dekel-Fudenberg procedure. Then given an arbitrary
cautious ordered epistemic model of $G,$ if $w\in \mathbb{CB}^{1}(LRAT),$
then $\sigma (w)\in S^{DF}$.\smallskip \newline
\textbf{Proof}. For each $n\in \mathbb{N},$ we use $S^{DFn}$ to denote the
set of strategy profiles surviving the first $n$ rounds of Dekel-Fudenberg
procedure. Let $\overline{M}=(W,\{R_{i}\}_{i\in N},\{\sigma _{i}\}_{i\in
N},\{\lambda _{i}\}_{i\in N})$ be a cautious ordered epistemic model of $G$
and $w\in W.$ We show that if $w\in \mathbb{CB}^{1}(LRAT),$ then $\sigma
(w)\in S^{DFn}$ for each $n\in \mathbb{N}.$ First, since $\overline{M}$ is
cautious, it can be seen that $\sigma (w)\in S^{DF1}$. Indeed, if there is
some $i\in N$ such that $\sigma _{i}(w)$ is eliminated in the first round of
Dekel-Fudenberg procedure, then there is some $r_{i}\in \Delta (S_{i}).$
Then it follows from Theorem 5.8.3 in Perea \cite{p12} (p.215, 221-226) $%
\sigma _{i}(w)$ cannot be optimal to any cauious belief, i.e., it cannot be
optimal on $\lambda _{i}(w),$ which is contradictory since $w\in LRAT_{i}$.

Now we show that $\sigma (w)\in S^{DF2}$. Suppose for some $i\in N$, $\sigma
_{i}(w)$ is eliminated in the second round of Dekel-Fudenberg procedure,
i.e., there exists $r_{i}\in \Delta (S_{i}^{DF1})$ such that $%
u_{i}(r_{i},s_{j})>u_{i}(s_{i},s_{j})$ for all $s_{j}\in S_{j}^{DF1}$. On
the other hand, since $w\in \mathbb{CB}^{1}(LRAT)$, $\sigma _{i}(w)$ is
optimal to $\lambda _{i}(w)(1).$ This implied that some strategies
supporting $\lambda _{i}(w)(1)$ has been eliminated in the first round.
However, since $w\in \mathbb{CB}^{1}(LRAT),$ it follows from the definition
that supp $\lambda _{i}(w)(1)\subseteq LRAT,$ which, from the argument
above, implies that all strategies $w^{\prime }\in $ supp $\lambda
_{i}(w)(1) $ should have survived the first round and $\sigma _{j}(w^{\prime
})$ stay in $S_{j}^{DF1},$ a contradiction.

Now suppose that $\sigma (w)\in S^{DF1}\cap ...\cap S^{DFn}$ but disappeared
in $S^{DFn+1}.$ This could happen only if some strategies supporting $%
\lambda _{i}(w)(1)$ had been eliminated in the $n$-th round, which is
because some strategies supporting that strategy in $\lambda _{i}(w)(1)$ in (%
$n-1$)-th round, etc. Finally this leads to the second and first rounds,
which, by the argument above, is impossible. Therefore $\sigma (w)\in
S^{DFn+1}$. //\smallskip \newline
\textbf{Proof of Theorem \ref{sec:las}.1: (1) }Since, by Proposition 5.2 in
Asheim and Dufwenberg \cite{ad03}, any strategy surviving Dekel-Fudenberg
procedure is permissible and vice versa, i.e., $S^{PER}=S^{DF},$ (1)
directly follows from Lemma \ref{sec:las}.1.\smallskip \newline
\textbf{(2) }Let $s\in S^{PER},$ that is, for each $i\in N,$ $s_{i}$ is
optimal to some type expressing common full belief in caution and primary
belief in rationality in a lexicographic epistemic model $%
M^{lex}=(T_{j},b_{j})_{j\in N}.$ We construct an ordered Kripke model $%
\overline{M}=(W,\{R_{i}\}_{i\in N},\{\sigma _{i}\}_{i\in N},\{\lambda
_{i}\}_{i\in N})$ based on $M^{lex}$ as follows:\smallskip \newline
(1) Let $W=T\times S,$ here $T=\Pi _{i\in N}T_{i};$\smallskip \newline
(2) for each $w=(t_{1},t_{2},s_{1},s_{2}),$ $\sigma _{i}(w)=s_{i};$%
\smallskip \newline
(3) Connectiong each state in $W$ according to $M^{lex},$ i.e., for each $%
w=(t_{1},t_{2},s_{1},s_{2}),w^{\prime }=(t_{1}^{\prime },t_{2}^{\prime
},s_{1}^{\prime },s_{2}^{\prime })\in T\times S,$ $w^{\prime }=\lambda
_{i}(w)(k)$ iff $t_{i}=t_{i}^{\prime },$ $s_{i}=s_{i}^{\prime },$ and $%
(s_{j}^{\prime },t_{j}^{\prime })$ is the $k$-th entry in $b_{i}(t_{i});$
mixed strategy-type pairs are defined in a similar way.\smallskip \newline
Without loss of generality, we can assume that each type in $M^{lex}$ is
cautious.\footnote{%
For a state that is not cautious we can extend it into a cautious one. See
Liu \cite{l18}.} It can be seen that $\overline{\mathcal{M}}$ is also
cautious, and there is $w\in W$ with $\sigma (w)=s$ and $w\in \mathbb{CB}%
^{1}(LRAT).$ //

\section{Ordered Kripke Model as the Limit of Probabilistic Kripke Models 
\label{sec:pkm}}

Though the ordered Kripke model is not the first framework combining
standarad Kripke model with an order on (a subset of) each $R_{i}(w)$ (cf.
Baltag and Smets \cite{bs06}, \cite{bs07}), here we are interested in how
such a model can be connected to the probabilistic Kripke model for games
introduced in Section \ref{sec:suy}. In this section we will first introduce
a probabilistic Kripke model with modified belief operators under which $%
\varepsilon $-perfect rationalizability can be characterized. Then we will
show that an ordered Kripke model can be seen as a \textquotedblleft
limit\textquotedblright\ of a sequence of probablistic Kripke models.

\subsection{Probabilistic belief and $\protect\varepsilon $-perfect
rationalizability}

In this subsection we give a survey on probabilitstic epistemic model (with
complete information) and the definition of $\varepsilon $-perfect
rationalizability. See Perea \cite{p12}, Chapter 2 for the detail of the
former. $\varepsilon $-permissible, which originates from Myerson \cite{m78}%
, is defined in a similar way as $\varepsilon $-proper rationalizability as
in Schuhmacher \cite{s99} and Perea and Roy \cite{ps17}. Consider a finite
2-person strategic form game $G=(N,\{S_{i}\}_{i\in N},\{u_{i}\}_{i\in N})$.
A finite \emph{probabilistic epistemic model} for $G$ is a tuple $%
M^{pro}=(T_{i},b_{i})_{i\in N}$ where\smallskip \newline
(a) $T_{i}$ is a finite set of types, and\smallskip \newline
(b) $b_{i}$ is a mapping that assigns to every $t_{i}\in T_{i}$ a
probability distribution over $\Delta (S_{j}\times T_{j}).$\smallskip 

For each $s_{i}\in S_{i}$ and $t_{i}\in T_{i},$ we define $%
u_{i}(s_{i},t_{i})=\Sigma _{(s_{j},t_{j})\in S_{j}\times
T_{j}}b_{i}(t_{i})(s_{j},t_{j})u_{i}(s_{i},s_{j}).$ $s_{i}$ is \emph{optimal}
(or \emph{rational}) for $t_{i}$ iff $u_{i}(s_{i},t_{i})\geq
u_{i}(s_{i}^{\prime },t_{i})$ for any $s_{i}^{\prime }\in S_{i}$. For each $%
s_{i},s_{i}^{\prime }\in S_{i}$ and $t_{i}\in T_{i},$ $s_{i}$ \emph{is
preferred to} $s_{i}^{\prime }$ under $t_{i}$ iff $%
u_{i}(s_{i},t_{i})>u_{i}(s_{i}^{\prime },t_{i}).$ Given $t_{i}\in T_{i},$
for each $(s_{j},t_{j})\in S_{j}\times T_{j},$ we say $t_{i}$ \emph{deems} $%
(s_{j},t_{j})$ \emph{possible} iff $b_{i}(t_{i})(s_{j},t_{j})>0.$ We say $%
t_{i}$ \emph{deems} $t_{j}\in T_{j}$ \emph{possible} iff $t_{i}$ deems $%
(s_{j},t_{j})$ possible for some $s_{j}\in S_{j}$. For each $t_{i}\in T_{i},$
we denote by $T_{j}(t_{i})$ the set of all $t_{j}$'s deemed possible by $%
t_{i}$.\smallskip \newline
\textbf{Definition \ref{sec:pkm}.1 (Caution)} Type $t_{i}\in T_{i}$ is \emph{%
cautious} iff for each $t_{j}\in T_{j}(t_{i})$ and each $s_{j}\in S_{j},$ $%
t_{i}$ deems $(s_{j},t_{j})$ possible.\smallskip \newline
\textbf{Definition \ref{sec:pkm}.2 (}$\varepsilon $\textbf{-perfect
trembling condition) }Type $t_{i}\in T_{i}$ satisfies $\varepsilon $\emph{%
-perfect trembling condition} iff for any $s_{j}\in S_{j}$ and $t_{j}\in
T_{j}(t_{i})$ such that $t_{i}$ deems $(s_{j},t_{j})$ possible, if $s_{j}$
is not optimal under $t_{j}$ then $b_{i}(t_{i})(s_{j},t_{j})\leq \varepsilon
.$\smallskip \newline
\textbf{Definition \ref{sec:pkm}.3 (Common full belief in a property) }Let $P
$ be an arbitrary property of probabilistic types.\smallskip \newline
(a) $t_{i}\in T_{i}$ \emph{expresses }$0$\emph{-fold full belief in} $P$ iff 
$t_{i}$ satisfies $P;$\smallskip \newline
(b) For each $n\in \mathbb{N},$ $t_{i}\in T_{i}$ \emph{expresses }$(n+1)$%
\emph{-fold full belief in} $P$ iff $t_{i}$ only deems possible $j$'s types
that express $n$-fold full belief in $P.$\smallskip \newline
$t_{i}$ \emph{expresses common full belief in} $P$ iff it expresses $n$-fold
full belief in $P$ for each $n\in \mathbb{N}.$\smallskip \newline
\textbf{Definition \ref{sec:pkm}.4 (}$\varepsilon $\textbf{-Perfect
rationalizability)}. Given a probabilistic epistemic model $%
M^{pro}=(T_{i},b_{i})_{i\in N}$ for a game $G=(N,\{S_{i}\}_{i\in
N},\{u_{i}\}_{i\in N})$, $s_{i}\in S_{i}$ is $\varepsilon $\textbf{-}\emph{%
permissible} iff it is optimal to some $t_{i}\in T_{i}$ which expresses
common full belief in caution and $\varepsilon $-perfect trembling
condition.\smallskip \newline
\textbf{Example \ref{sec:pkm}.1}. Consider the following game $G$ (from
Myerson \cite{m78}):%
\begin{equation*}
\begin{tabular}{|l|l|l|}
\hline
$u_{1}\backslash u_{2}$ & $C$ & $D$ \\ \hline
$A$ & $1,1$ & $0,0$ \\ \hline
$B$ & $0,0$ & $0,0$ \\ \hline
\end{tabular}%
\end{equation*}%
and $M^{pro}=(T_{i},b_{i})_{i\in N}$ for $G$ where $T_{1}=\{t_{1}\},$ $%
T_{2}=\{t_{2}\},$ and%
\begin{equation*}
b_{1}(t_{1})=(1-\varepsilon )(C,t_{2})+\varepsilon (D,t_{2}),\text{ }%
b_{2}(t_{2})=(1-\varepsilon )(A,t_{1})+\varepsilon (B,t_{1}),
\end{equation*}%
where $\varepsilon \in (0,1).$ It can be seen that $A$ is $\varepsilon $%
-permissible since it is optimal to $t_{1}$ which expresses common full
belief in caution and $\varepsilon $-perfect trembling condition.\smallskip 

Originally, by the definition in Myerson \cite{m78}, perfect equilibrium is
the limit of $\varepsilon $-perfect equilibrium. Though permissibility is
the concepts in epistemic game theory and is defined in a different manner,
it still holds tha permissibility is the limit of $\varepsilon $%
-permissibility. See Schuhmacher \cite{s99}.

\subsection{Characterizing $\protect\varepsilon $-permissibility in
probabilistic Kripke model}

In this subsection, we show how to use probabilistic Kripke model to
describe $\varepsilon $-permissibility. Let $G=(N,\{S_{i}\}_{i\in
N},\{u_{i}\}_{i\in N})$ be a 2-person strategic form game and $\mathcal{M}%
=(W,\{R_{i}\}_{i\in N},\{p_{i}\}_{i\in N},\{\sigma _{i}\}_{i\in N})$ a
probabilistic Kripke model for $G.$ We give the following
definitions\smallskip \newline
\textbf{Definition \ref{sec:pkm}.5 (Caution)}. $\mathcal{M}$ is \emph{%
cautious} at $w\in W$ for $i\in N$ iff for any $s_{j}\in S_{j}$ ($j\neq i$),
there exists $w^{\prime }\in R_{i}(w)$ satisfying $p_{i}(w)(w^{\prime })>0$
and $\sigma _{j}(w^{\prime })=s_{j}$. We say $\mathcal{M}$ is \emph{cautious}
iff $\mathcal{M}$ is cautious at every $w\in W$ for each $i\in N$.\smallskip 
\newline
\textbf{Definition \ref{sec:pkm}.6 (}$\varepsilon $\textbf{-perfect
trembling condition)}. $\mathcal{M}$ satisfies $\varepsilon $\emph{-perfect
trembling condition} at $w\in W$ for $i\in N$ iff for each $w^{\prime }\in
R_{i}(w),$ if $\sigma _{i}(w^{\prime })$ (i.e., $\sigma _{i}(w)$) is not
optimal to $\sigma _{j}(w^{\prime }),$ then $p_{i}(w)(w^{\prime })\leq
\varepsilon $.\smallskip 

The above two concepts are illustrated in the following example.\smallskip 
\newline
\textbf{Example \ref{sec:pkm}.1}. Consider the game $G$ in Example \ref%
{sec:suy}.1:%
\begin{equation*}
\begin{tabular}{|l|l|l|}
\hline
$u_{1}\backslash u_{2}$ & $C$ & $D$ \\ \hline
$A$ & $1,1$ & $0,0$ \\ \hline
$B$ & $0,0$ & $0,0$ \\ \hline
\end{tabular}%
\end{equation*}%
and a probabilistic Kripke model depicted as in Figure 2. 

\begin{figure}[h]
\centering\includegraphics[width=0.9\linewidth]{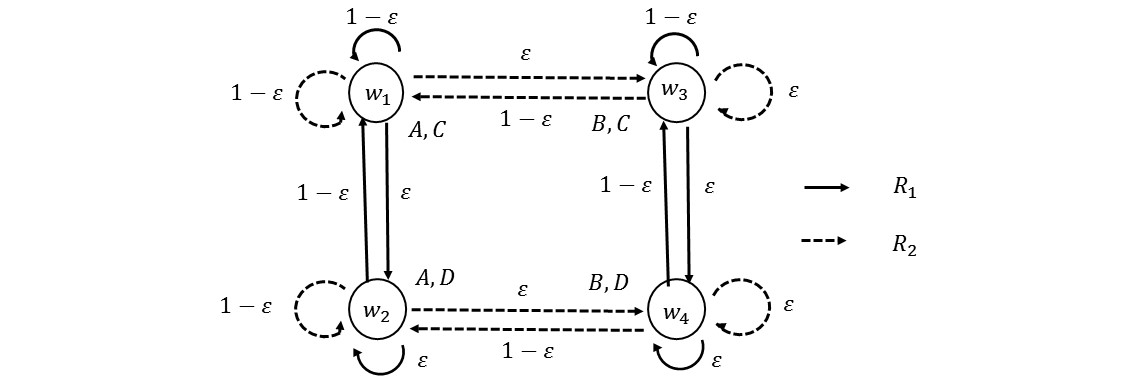}
\caption{An probabilistic Kripke model for $G$}
\end{figure}

It can be seen that $%
RAT_{1}=\{w_{1},w_{2}\}$, $RAT_{2}=\{w_{1},w_{3}\},$ and $RAT=RAT_{1}\cap
RAT_{2}=\{w_{1}\}.$ Since $\sigma (w_{1})=(A,C)$ is a pair of $\varepsilon $%
-perfect rationalizable strategies, $RAT$ can still be used in this
framework for the first-order. Now the problem is how to describe higher
orders, i.e., interpersonal belief and common full belief in this framework. 
$\mathbb{B}_{i}$ and $\mathbb{CB}$ in standard probabilistic model do not
work here since $\mathbb{B}_{i}(RAT)=\mathbb{CB(}RAT)=\emptyset ,$ while we
want to keep $w_{1}.$ Here we provide an approach. Let $\mathcal{M}%
=(W,\{R_{i}\}_{i\in N},\{p_{i}\}_{i\in N},\{\sigma _{i}\}_{i\in N},)$ a
probabilistic Kripke model for $G$ satisfying caution and $\varepsilon $%
-perfect trembling condition. For each $i\in N,$ we define $%
R_{i}^{>\varepsilon }=\{w^{\prime }\in R_{i}(w):p_{i}(w)(w^{\prime
})>\varepsilon \}.$ An\emph{\ upper }$\varepsilon $ \emph{semantic belief
operator} is a function $\mathbb{B}_{i}^{>\varepsilon }:2^{W}\rightarrow
2^{W}$ such that for each $E\subseteq W,$%
\begin{equation}
\mathbb{B}_{i}^{>\varepsilon }(E)=\{w\in W:R_{i}^{>\varepsilon }(w)\subseteq
E\}.  \label{oi0}
\end{equation}%
An \emph{upper }$\varepsilon $ \emph{semantic common belief operator} is a
function $\mathbb{CB}^{>\varepsilon }:2^{W}\rightarrow 2^{W}$ such that for
each $E\subseteq W,$%
\begin{equation}
\mathbb{CB}^{>\varepsilon }(E)=\{w\in W:\cup _{i\in N}R_{i}^{>\varepsilon
}(w)\subseteq E\}.  \label{oi9}
\end{equation}%
When $\varepsilon <\frac{1}{2},$ it can be seen that in Example \ref{sec:pkm}%
.1, $\mathbb{CB}^{>\varepsilon }(RAT)=\{w_{1}\}.$\smallskip

In general, we have the following statement.\smallskip \newline
\textbf{Theorem \ref{sec:pkm}.1 (Characterizing }$\varepsilon $\textbf{%
-perfect rationalizability)}. Let $G=(N,\{S_{i}\}_{i\in N},\{u_{i}\}_{i\in
N})$ be a 2-person strategic form game, $\varepsilon <\frac{1}{2},\footnote{%
It should be noted that $\varepsilon <\frac{1}{2}$ makes sure that $%
R_{i}^{>\varepsilon }(w)\neq R_{i}(w),$ though from the viewpoint of
convergence/limit this is just a technical requirement.}$ and $%
S^{\varepsilon PER}\subseteq S$ be the set of $\varepsilon $-permissible
strategy profiles. Then\smallskip \newline
\textbf{(1)} given an arbitrary probabilistic Kripke model of $G$ satisfying
caution and $\varepsilon $-perfect trembling condition, if $w\in \mathbb{CB}%
^{>\varepsilon }(RAT),$ then $\sigma (w)\in S^{\varepsilon PER}$,
and\smallskip \newline
\textbf{(2)} for each $s\in S^{\varepsilon PER},$ there exists a
probabilistic Kripke model of $G$ satisfying caution and $\varepsilon $%
-perfect trembling condition such that $\sigma (w)=s$ and $w\in \mathbb{CB}%
^{>\varepsilon }(RAT).$\smallskip \newline
\textbf{Proof}. (2) can be proved in a similar way as Theorem \ref{sec:las}%
.1. Here we only prove (1). Since there is no algorithm like Dekel-Fudenberg
procedure that can screen out $\varepsilon $-permissibility, we show how to
construct a type which expresses common full belief in caution and $%
\varepsilon $-perfect trembling condition. Let $\mathcal{M}%
=(W,\{R_{i}\}_{i\in N},\{p_{i}\}_{i\in N},\{\sigma _{i}\}_{i\in N})$ be a
probabilistic Kripke model for $G$ satisfying caution and $\varepsilon $%
-perfect trembling condition and $w\in \mathbb{CB}^{>\varepsilon }(RAT)$.
For $i\in N,$ we define a partition $\mathbb{E}_{i}=\{E_{i1},...,E_{i\ell
_{i}}\}$ of $W$ and satisfies that for each $w^{\prime },w^{\prime \prime
}\in W,$ $w^{\prime }$ and $w^{\prime \prime }$ belong to the same
equivalent class $E_{i}$ if and only if $R_{i}(w^{\prime })=R_{i}(w^{\prime
\prime })$ and $p_{i}(w^{\prime })=p_{i}(w^{\prime \prime }).$ For each $%
E_{i}\in \mathbb{E}_{i}$ we assign a symble $t_{i}(E_{i}).$ Without loss of
generality, we can assume that for each $s_{i}\in S_{i}$ and each $E_{ik},$
there is some $w^{\prime }\in E_{ik}$ such that $\sigma _{i}(w^{\prime
})=s_{i}.\footnote{%
This corresponds to caution for probabilistic epistemic model $M^{pro}.$ It
should be noted that even this condition is not satisfied, we can construct
\textquotedblleft dummies\textquotedblright\ to make this condition
satisfied without hurt the model.}$ Let $T_{i}=\{t_{i}(E_{i})\}_{E_{i}\in 
\mathbb{E}_{i}}$, and define $b_{i}(t_{i}(E_{i}))$ with the same probability
as $p_{i}(w^{\prime })$, where $w^{\prime }\in E_{i},$ and the corresponding 
$t_{j}(E_{j})$. It can be seen that $b_{i}(t_{i}(E_{i}))$ is well-defined
since every state in one $E_{i}$ has identified distributions. It can be
seen straightforwardly that $\sigma (w)\in S^{\varepsilon PER}$ since for
each $i\in N,$ $\sigma _{i}(w)$ is optimal to the type corresponding to $w$
which expresses common full belief in caution and $\varepsilon $-perfect
trembling condition since $\mathcal{M}$ satisfies caution and $\varepsilon $%
-perfect trembling condition. //

\subsection{Probabilistic Kripke models converge to ordered Kripke model}

In this subsection we show that any ordered Kripke model is the limit of a
sequence of probabilistic Kripke models. This can be intuitively seen by
comparing the two Kripke models in Example \ref{sec:las}.1 and \ref{sec:pkm}%
.1: Indeed, Figure 2 can be obtained by replacing $%
1-\varepsilon $ with $1$ and $\varepsilon $ with $2$ in Figure 1. Also, this can be seen from that perfect rationalizability characterized
by the former is the limit of a sequence of $\varepsilon $-perfect
rationalizabilities characterized by the later. In this section we show how
to formulate this idea.

Let $G=(N,\{S_{i}\}_{i\in N},\{u_{i}\}_{i\in N})$ be a 2-person strategic
form game and $\overline{\mathcal{M}}=(W,\{R_{i}\}_{i\in N},$ $\{\sigma
_{i}\}_{i\in N},\{\lambda _{i}\}_{i\in N})$ be an ordered Kripke model of $G.
$ Without loss of generality, we assume that $\overline{\mathcal{M}}$
satisfies the following two conditions:\smallskip \newline
\textbf{(Disjoint supports)} For each $i\in N$, $w\in W,$ and $k,k^{\prime
}\in \mathcal{D}(\lambda _{i}(w)),$ supp $\lambda _{i}(w)(k)\cap $ supp $%
\lambda _{i}(w)(k^{\prime })\neq \emptyset $ if and only if $k\neq k^{\prime
};\footnote{%
This condition is adopted in some papers such as Blume et al. \cite{bbd91a}, 
\cite{bbd91b} while is not required in some others such as the standard
textbook of Perea \cite{p12}. Technically, this condition is not necessary
in characterizing rationalizabilities. Here we use it out of simplification.}
$\smallskip \newline
\textbf{(Surjection)} For each $w\in W$ and each $w^{\prime }\in R_{i}(w),$
there is some $k\in \mathcal{D}(\lambda _{i}(w))$ such that $\lambda
_{i}(w)(k)(w^{\prime })>0.\footnote{%
Surjection is different from caution. Caution requires each strategy of the
opponent should appear in the range. When there are multiple states in $%
R_{i}(w)$ which are assigned the same strategy, to be cautious only means
that at least one of those state should appear in the range, while
surjection requires that each of these states should appear. On the other
hand, it does not mean that surjection implies caution since surjection has
nothing with strategies assigned to each state in $R_{i}(w)$. In finite
models, when surjection is not satisfied, we can faithfully extend each
model into one which satisfies surjection without hurting $LRAT$ and $%
\mathbb{CB}(LRAT)$.}$\smallskip 

Let $\varepsilon \in (0,1)$. Consider a probabilistic Kripke model $\mathcal{%
M}^{\varepsilon }=(W^{\varepsilon },\{R_{i}^{\varepsilon }\}_{i\in
N},\{p_{i}^{\varepsilon }\}_{i\in N},\{\sigma _{i}^{\varepsilon }\}_{i\in N})
$ of $G$ satisfying\smallskip \newline
(a) $W^{\varepsilon }=W,$ $R_{i}^{\varepsilon }=R_{i}$ and $\sigma
_{i}^{\varepsilon }=\sigma _{i}$ for each $i\in N;$\smallskip \newline
(b) for each $i\in N,$ $w\in W$, and $w^{\prime },w^{\prime \prime }\in $
supp $\lambda _{i}(w)(k)$ for some $k\in \mathcal{D}(\lambda _{i}(w)),$ it
is satisfied that $p_{i}^{\varepsilon }(w)(w^{\prime })/p_{i}^{\varepsilon
}(w)(w^{\prime \prime })=\lambda _{i}(w)(k)(w^{\prime })/\lambda
_{i}(w)(k)(w^{\prime \prime })$;\smallskip \newline
(c) for each $i\in N,$ $w\in W$, and $w^{\prime }\in R_{i}(w),$ $%
0<p_{i}^{\varepsilon }(w)(w^{\prime })\leq \varepsilon $ if $\lambda
_{i}(w)(1)(w^{\prime })=0.$\smallskip \newline
It can be seen that when $\varepsilon $ is small enough, such $\mathcal{M}%
^{\varepsilon }$ exists (not unique). Let $\{\varepsilon _{n}\}_{n\in 
\mathbb{N}}$ be a sequence in $(0,1)$ that converges to $0$ such that for
each $\varepsilon _{n},$ there is some probabilistic Kripke model satisfying
(a) - (c). We choose an arbitrary $\mathcal{M}^{\varepsilon _{n}}$
satisfying (a) - (c) for each $\varepsilon _{n}.$ It can be seen that the
sequence $\{\mathcal{M}^{\varepsilon _{n}}\}_{n\in \mathbb{N}}$
\textquotedblleft converges\textquotedblright\ to $\overline{\mathcal{M}}$
in the sense that\smallskip \newline
(1) for each $i\in N,$ $w\in W$, and $w^{\prime }\in R_{i}(w)$ such that $%
p_{i}^{\varepsilon _{n}}(w)(w^{\prime })\rightarrow 0,$ $w^{\prime }$ does
not appear in $\lambda _{i}(w)(1);$\smallskip \newline
(2) for each $i\in N,$ $w\in W$, and $w^{\prime }\in R_{i}(w)$ such that $%
p_{i}^{\varepsilon _{n}}(w)(w^{\prime })\nrightarrow 0,$ $w^{\prime }\in $
supp $\lambda _{i}(w)(1)$ and $p_{i}^{\varepsilon _{n}}(w)(w^{\prime
})\rightarrow \lambda _{i}(w)(1)(w^{\prime });$\smallskip \newline
(3) The convergence is propotional within each level of $\lambda _{i}(w).%
\footnote{%
To dealing those \textquotedblleft irrational\textquotedblright\ choice
which is assigned in probability $0$ in any rational belief is one of the
motivation for the introduce of lexicographic belief and studies from
conditional probability. See Blume et al. \cite{bbd91a}, \cite{bbd91b},
Brandenburger et al. \cite{bfk07}, Halpern \cite{h10}.}$\smallskip \newline
More formally, this convergence can be seen from the rationalizabilities
they characterize. We have the following statement.\smallskip \newline
\textbf{Theorem \ref{sec:pkm}.2 (Probabilistic models converge to ordered
model)}. Let $G$ be a 2-person strategic game, $\overline{\mathcal{M}}$ a
cautious ordered Kripke model of $G$ satisfying disjoint supports and
surjection$,$ $\{\varepsilon _{n}\}_{n\in \mathbb{N}}$ a sequence in $(0,1)$
converging to $0,$ and $\{\mathcal{M}^{\varepsilon _{n}}\}_{n\in \mathbb{N}}$
be a sequence of probabilistic model of $G$ satisfying condition (a) - (c)
above for each $\varepsilon _{n}.$ Then each $w$ which is commonly believed
to be lexicographically rational in $\overline{\mathcal{M}}$ is commonly
believed to be $\varepsilon _{n}$-upper rational in $\mathcal{M}%
^{\varepsilon _{n}}$ for each $\varepsilon _{n},$ i.e., $\mathbb{CB}%
^{1}(LRAT)=\cup _{M\in \mathbb{N}}\cap _{n>M}\mathbb{CB}^{>\varepsilon
_{n}}(RAT_{\varepsilon _{n}})$.\smallskip \newline
\textbf{Proof}. ($\subseteq $) Let $w\in \mathbb{CB}^{1}(LRAT).$ It follows
from Theorem \ref{sec:las}.1 that $\sigma (w)\in S^{PER},$ that is, there is
a lexicographic model $(\Theta _{i},\beta _{i})_{i\in N}$ such for each $%
i\in N$, $\sigma _{i}(w)$ is optimal to some $\theta _{i}\in \Theta _{i}$
which expresses common full belief in caution and primary belief in
rationality. Based on each $\mathcal{M}^{\varepsilon _{n}},$ $\theta _{i}$
can be accordingly translated into a state $t_{i}^{\varepsilon _{n}}$
\textquotedblleft starting\textquotedblright\ from $w$ in probabilistic
model. Since $\sigma _{i}(w)$ is optimal to some $\theta _{i},$ when $%
\varepsilon _{n}$ is small enough, $\sigma _{i}(w)$ is optimal to $%
t_{i}^{\varepsilon _{n}},$ and $t_{i}^{\varepsilon _{n}}$ expresses common
full belief on caution and $\varepsilon _{n}$-perfect trembling condition.
This argument holds for each $i\in N.$ Then by Theorem \ref{sec:pkm}.1 $%
\sigma (w)\in \mathbb{CB}^{>\varepsilon _{n}}(RAT_{\varepsilon _{n}}),$ and
consequently $w\in \cup _{M\in \mathbb{N}}\cap _{n>M}\mathbb{CB}%
^{>\varepsilon _{n}}(RAT_{\varepsilon _{n}})$.

($\supseteq $) Let $w\in \cup _{M\in \mathbb{N}}\cap _{n>M}\mathbb{CB}%
^{>\varepsilon _{n}}(RAT_{\varepsilon _{n}})$, that is, for some $M\in 
\mathbb{N},$ $w\in \mathbb{CB}^{>\varepsilon _{n}}(RAT_{\varepsilon _{n}})$
for all $n\geq M.$ Since $\varepsilon _{n}\rightarrow 0,$ it follows that
for each $i\in N,$ $\sigma _{i}(w)$ is optimal on $p_{i}^{\varepsilon
_{n}}(w)$ for infinitely small $\varepsilon _{n}.$ Since each $\mathcal{M}%
^{\varepsilon _{n}}$ keeps the propotion between states within each level of 
$\lambda _{i}(w),$ this implies that $\sigma _{i}(w)$ is optimal to $\lambda
_{i}(w).$ Also, the state in the probabilistic model of $G$ corresponding to 
$\mathcal{M}^{\varepsilon _{n}}$ supporting $\sigma _{i}(w)$ expresses
common full belief in caution and $\varepsilon $-perfect trembling
condition. Since $\overline{\mathcal{M}}$ is surjective, it follows that the
corresponding type in the lexicographic model for $\overline{\mathcal{M}}$
expresses common full belief in caution and primary belief in rationality.
Therefore $\sigma _{i}(w)$ is perfect rationalizable in $\overline{\mathcal{M%
}}$. Since this argument holds for all $i\in N,$ it follows that $w\in 
\mathbb{CB}^{1}(LRAT).$ //

\section{Concluding Remarks \label{sec:cr}}

\subsection{Convergence and proper rationalizability}

In Section \ref{sec:las}, we characterized permissibility by ordered Kripke
model. Though it is desirable to characterize other rationalizability
concepts, e.g., proper rationalizability (Asheim \cite{a01}), in the ordered
Kripke model, we think it is difficult, if not impossible. The reason is
that in this framework, the difference between perfect and proper
rationalizabilities is at what kind of order $\lambda _{i}(w)$ gives on $%
R_{i}(w),$ which more relies on the interpretation than on the structure. In
other words, by changing the order on accessable states we can characterize
proper rationalizability; but this is attributed to the interpretation we
give to each state, not to any structural properties of the Kripke frame $%
(W,\{R_{i}\}_{i\in N})$ like seriality or transitivity.

On the other hand, using the approach introduced in Section \ref{sec:pkm}.3,
proper rationalizability can be discussed as the limit of probabilistic
Kripke models. Let $G$ be a 2-person strategic form game and a cautious
ordered Kripke model $\overline{\mathcal{M}}$ satisfies a condition parallel
to \textquotedblleft respecting the opponent's preferences\textquotedblright
. For $\varepsilon >0,$ consider $\mathcal{M}^{\varepsilon }=(W^{\varepsilon
},\{R_{i}^{\varepsilon }\}_{i\in N},\{p_{i}^{\varepsilon }\}_{i\in
N},\{\sigma _{i}^{\varepsilon }\}_{i\in N})$ a probabilistic Kripke model of 
$G$ satisfying conditions (a), (b) in Section \ref{sec:pkm}.3 and (c$%
^{\prime }$) for each $i\in N,$ $w\in W$, and $w^{\prime },w^{\prime \prime
}\in R_{i}(w),$ $0<p_{i}^{\varepsilon }(w)(w^{\prime })\leq \varepsilon
p_{i}^{\varepsilon }(w)(w^{\prime \prime })$ if $\lambda _{i}(w)(k^{\prime
})(w^{\prime })>0$, $\lambda _{i}(w)(k^{\prime \prime })(w^{\prime \prime
})>0$, and $k^{\prime \prime }>k^{\prime }.$ It can be seen that (1) each
probabilistic Kripke model satisfying (a), (b), and (c$^{\prime }$)
characterizes some $\varepsilon $-perfect rationalizable strategies; (2) $%
\overline{\mathcal{M}}$ is the limit of a sequence $\{\mathcal{M}%
^{\varepsilon _{n}}\}_{n\in \mathbb{N}}$ satisfying (a), (b), and (c$%
^{\prime }$); and (3) the perfect rationalizable strategies characterized in 
$\overline{\mathcal{M}}$ is limits of $\varepsilon _{n}$-perfect
rationalizable strategies characterized by $\{\mathcal{M}^{\varepsilon
_{n}}\}_{n\in \mathbb{N}}$.

\subsection{Syntactical system}

In this paper we have defined the ordered Kripke model to capture the
concept of rationality under lexicographic belief hierarchy by a semantical
approach. It is wondered that whether there exists a syntactic approach
corresponding to that semantic framework, like the one developed in Bonanno 
\cite{b08} for the standard Kripke model for games. A critical property for
that syntactic system, if exists, is that the change of the criterion for
truth value from the first order to higher orders in the hierarchy, that is,
in the first order we need (at most) to check every accessible state, while
in the second order $\mathbb{B}_{i}^{1}$ we need only to check the first
level states, etc. Works are expected in this direction.

\end{document}